\documentclass[11pt,twoside]{article}
\usepackage{rr-asp2010}
\usepackage[breaklinks=true]{hyperref} 

\def\specchar#1{\uppercase{#1}} 
\def\rmit#1{{\rm #1}}              
\def\Halpha{\mbox{H\hspace{0.1ex}$\alpha$}}
\def\CaII{\mbox{Ca\,\specchar{ii}}} 
\def\CaIIH{\mbox{Ca\,\specchar{ii}\,\,H}}
\def\MgII{\mbox{Mg\,\specchar{ii}}} 
\def\hk{\mbox{h\,\&\,k}}
\def\is{\!=\!}  
\def\cf{cf.}                       
\def\ie{\rmit{i.e.,}}              
\def\kms{\hbox{km$\;$s$^{-1}$}}
\bibpunct{(}{)}{;}{a}{}{,}    
\def\citeads#1{\href{http://adsabs.harvard.edu/abs/#1}{\citealp{#1}}}
\def\citetads#1{\href{http://adsabs.harvard.edu/abs/#1}{\citet{#1}}}

\resetcounters
\begin{document}

\title{Twists to Solar Spicules}
\author{Robert J. Rutten$^{1,2}$
\affil{$^1$Lingezicht Astrophysics, 
       't Oosteneind 9, 4158 CA Deil, The Netherlands}}
\affil{$^2$Institute of Theoretical Astrophysics,
         University of Oslo,
         P.O. Box 1029, Blindern, N--0315 Oslo, Norway}

\begin{abstract}
  Type-II solar spicules appear as long, thin, highly dynamic strands
  of field-tied matter that feed significant mass and energy to the
  corona and solar wind.  A recent result is that they exhibit
  torsional Alfv\'en waves in addition to accelerating outflows and
  swaying motions due to transverse Alfv\'enic waves.  I summarize this
  finding and then re-interpret older observations in its light: the
  striking similarity of near-limb scenes in the outer blue and red
  wings of \Halpha, and the tilts of absorption lines with respect to
  emission lines in eclipse spectra taken in 1973.
\end{abstract}

\section{Utrecht solar physics}

I spent 50 years in Utrecht astronomy, arriving as a student in the
autumn of 1961 and cleaning out my desk at the near-defunct
Sterrekundig Instituut Utrecht (SIU) in the autumn of 2011\footnote{
  My last act at the SIU was to scan the lecture notes and practicals
  (in Dutch) of M.G.J.~Minnaert who in 1961--1963 gave his
  wide-ranging undergraduate courses (one year planets, one year
  stars) for the last time.  I posted them on my website (Google ``Rob
  Rutten'').  There I also describe the SIU closure and list
  the nearly 60 Utrecht solar physics PhDs.}.  Mostly in solar
physics, making me expect an invitation to review Utrecht solar
physics for you.  When it didn't come I offered this contributed talk
on recent results, but I place these here in earlier Utrecht context.
And let me summarily add that in the hands of Minnaert, de Jager and
Zwaan with their pupils Utrecht solar physics was a major player in
the field.  Utrecht (meaning Dutch) solar physics is now dead, but its
legacy lives on abroad\footnote{ Utrecht University alumni employed
  elsewhere who are active in solar physics and collectively embody
  the great name that Utrecht University had in this field but
  discarded, in PhD order: Henk Spruit (M\"unchen, Germany), Aad van
  Ballegooijen (Cambridge, USA), Piet Martens (Bozeman, USA), Karel
  Schrijver (Palo Alto, USA), Paul Hick (San Diego, USA), Han
  Uitenbroek (Sunspot, USA), Jo Bruls (Freiburg, Germany), Martin
  Volwerk (Graz, Austria), Kostas Tziotziou (Athens, Greece), Luc
  Rouppe van der Voort (Oslo, Norway), Michiel van Noort (Lindau,
  Germany), Alfred de Wijn (Boulder, USA), Jorrit Leenaarts (Oslo,
  Norway), Nikola Vitas (La Laguna, Spain), Catherine Fischer
  (Noordwijk, ESA), Gregal Vissers (Oslo, Norway), Tijmen van Wettum
  (Freiburg, Germany).  Plus active-pensioner 
  alumni Kees de Jager (Texel, The Netherlands), Jacques Beckers
  (Scottsdale, USA), myself.}.

\section{Chromosphere research at Utrecht and elsewhere}
Soon after I started at Utrecht Beckers defended his famous thesis on
the fine structure of the chromosphere
(\citeads{1964PhDT........83B})\footnote{ 
  A Utrecht thesis with Minnaert as supervisor, but Beckers had done the
  work in Australia and moved to the Sacramento Peak Observatory.
  Recently he scanned the full text; I put it on ADS.  Regrettably,
  the other Utrecht theses have not been scanned so far.}.  At Utrecht
the efforts in chromosphere research at that time consisted of
Houtgast's eclipse expeditions to collect high-dispersion flash
spectrum photography and ``Utrecht reference model'' building by De
Jager with Heintze and Hubenet.  Houtgast took me along to Greece and
Brazil in 1966, to Mexico in 1970; I wrote my thesis on partial
redistribution in line formation on data from the third expedition,
with Zwaan as supervisor (I am proud that I was his first) and on the
same topic as Houtgast's own famous thesis
(\citeads{1942QB551.H68......}). 
By then the basics of chromospheric NLTE-PRD line formation were well
established, a landmark of early computational astrophysics masterly
canonized in Mihalas (\citeyear{1970stat.book.....M},
\citeyear{1978stat.book.....M}).  Stuff I still teach with
pleasure.

During the following three decades not much happened in chromosphere
physics worldwide.  In retrospect the field was on hold until
computers matured.  The chromosphere is the most difficult solar
regime, far beyond snapshot observation and analytic modeling; it
needs fast computation.  In groundbased observing to permit real-time
adaptive optics and large-volume post-detection wavefront restoration,
in space-based observing to cope with high-cadence image streams, and
in modeling to permit 3D(t) MHD simulations with realistic radiative
transfer.  Data dissection and visualization are major issues in each
endeavor.  Eventually they all advanced enough to turn the
chromosphere from a non-doable topic into a hot topic.  See
\citetads{2011arXiv1110.6606R} 
for the latest review\footnote{ Since then there is significant
  progress in our understanding of the formation of \Halpha, the
  quintessential chromosphere diagnostic
  (\citeads{2012A&A...540A..86R}; 
  \citeads{2012ApJ...749..136L}). 
  I don't have space here to explain it but do in my (Google for it)
  ``Graphical introduction to NLTE chromospheric line formation''.}.

Utrecht solar physics became part of this chromosphere revival when
Hammerschlag's Dutch Open Telescope (DOT, \url{http://www.dot.iac.es})
started taking \Halpha\ movies. 
Utrecht left it in September 2011 when Leenaarts took his Veni grant
on \Halpha\ simulation to Oslo (the best leave first).
 
\section{New twist to spicules}

\altsubsubsection*{Dynamic fibrils$\,/\,$spicules-I and 
  straws$\,/\,$spicules-II$\,/\,$RBEs}
Still at the Sacramento Peak Observatory, Beckers wrote two
authoritative reviews on solar spicules
(\citeauthor{1968SoPh....3..367B} 
\citeyear{1968SoPh....3..367B}, 
\citeyear{1972ARA&A..10...73B}), 
a topic that thereafter suffered from utter lack of progress until
dynamic fibrils were identified and explained by
\citetads{2006ApJ...647L..73H} 
and \citetads{2007ApJ...655..624D}. 
These are slanted-field wave guides in which $p$-mode-driven
magneto-acoustic shock waves push matter up repetitively.  Just off
the limb they make up the dense forest of what we now call type-I
spicules.

Type-II spicules extend much further up, as very thin, very dynamic
strands.  They were first glimpsed by me as long thin ``straws'' in
near-limb \CaIIH\ movies from the DOT
(\citeads{2006ASPC..354..276R}), 
then observed in off-limb \CaIIH\ movies from Hinode by
\citetads{2007PASJ...59S.655D}, 
and subsequently identified on the disk as blue-wing absorption
features in \Halpha\ (Rapid Blue Excursions, RBEs) by
\citetads{2009ApJ...705..272R}. 
\citetads{2007Sci...318.1574D} 
identified their large off-limb swaying as upward propagating
transverse Alfv\'enic waves.  They are now thought to be important
contributors to coronal heating and mass loading
(\citeads{2009ApJ...701L...1D}), 
shooting blobs of matter up as bullets that are also diagnosed in
extreme-ultraviolet imaging
(\citeads{2011Sci...331...55D}) 
and spectroscopy
(\citeads{2011ApJ...738...18T}). 

\altsubsubsection*{Torsion waves next to outflows and swaying}
A new aspect, namely twist, of type-II spicules emerged during the
past year.  This finding is being published elsewhere
(\citeads{2012arXiv1205.5006D}). 
It was inspired by old near-limb DOT \Halpha-sampling observations
that in the far wings showed only upright fibrils, rooted in network,
much like straws and off-limb type-II spicules, whereas at line center
a dense carpet of long horizontally spread-out fibrils covering cell
interiors is seen (Fig.~7 of
\citeads{2007ASPC..368...27R}; 
Fig.~\ref{fig:DOT-nine} here).

This outer-wing scene is remarkably similar in the blue and red wings.
At the time this similarity was puzzling because spicules were then
thought to be slanted extending and retracting jets.  If the
corresponding Dopplershifts would define their outer-wing appearance
one would expect larger Dopplershift plus spatial foreshortening for
spicules slanted towards the observer.  A.G.~de Wijn then suggested
that the similarity might come from huge thermal broadening which
would produce identical outer-wing appearance, but we couldn't verify
this idea because the DOT speckle-burst profile sampling took too long
to compare red and blue outer-wing scenes sufficiently fast.

Our new analysis uses much higher-cadence spectral sampling with the
Swedish 1-m Solar Telescope (SST).  These data show that the
outer-wing scenes have indeed large similarity but are not identical.
Often the similarity between the two wings improves when comparing
them at a brief time delay.  B.~De~Pontieu then suggested that the
similarity results from twisting motions with rapid sign changes
giving transverse Dopplershifts.  Such spicule twists were earlier
suggested by \citetads{2008ASPC..397...27S} 
but without Doppler discrimination.  The clincher SST observation was
that these motions produce tilted spicular Dopplershift signature in
off-limb \CaIIH\ spectra and near-limb \CaII\,854.2\,nm imaging
spectroscopy.  The tilts change fast.  A Monte Carlo simulation served
to establish that their complex time-dependent spectral and spatial
behavior results from combining outward accelerating upflows with both
transverse and torsional Alfv\'en waves that also propagate upwards,
with much phase and periodicity mixing. See 
\citetads{2012arXiv1205.5006D} 
for more detail.

\articlefigure[width=11.4cm]{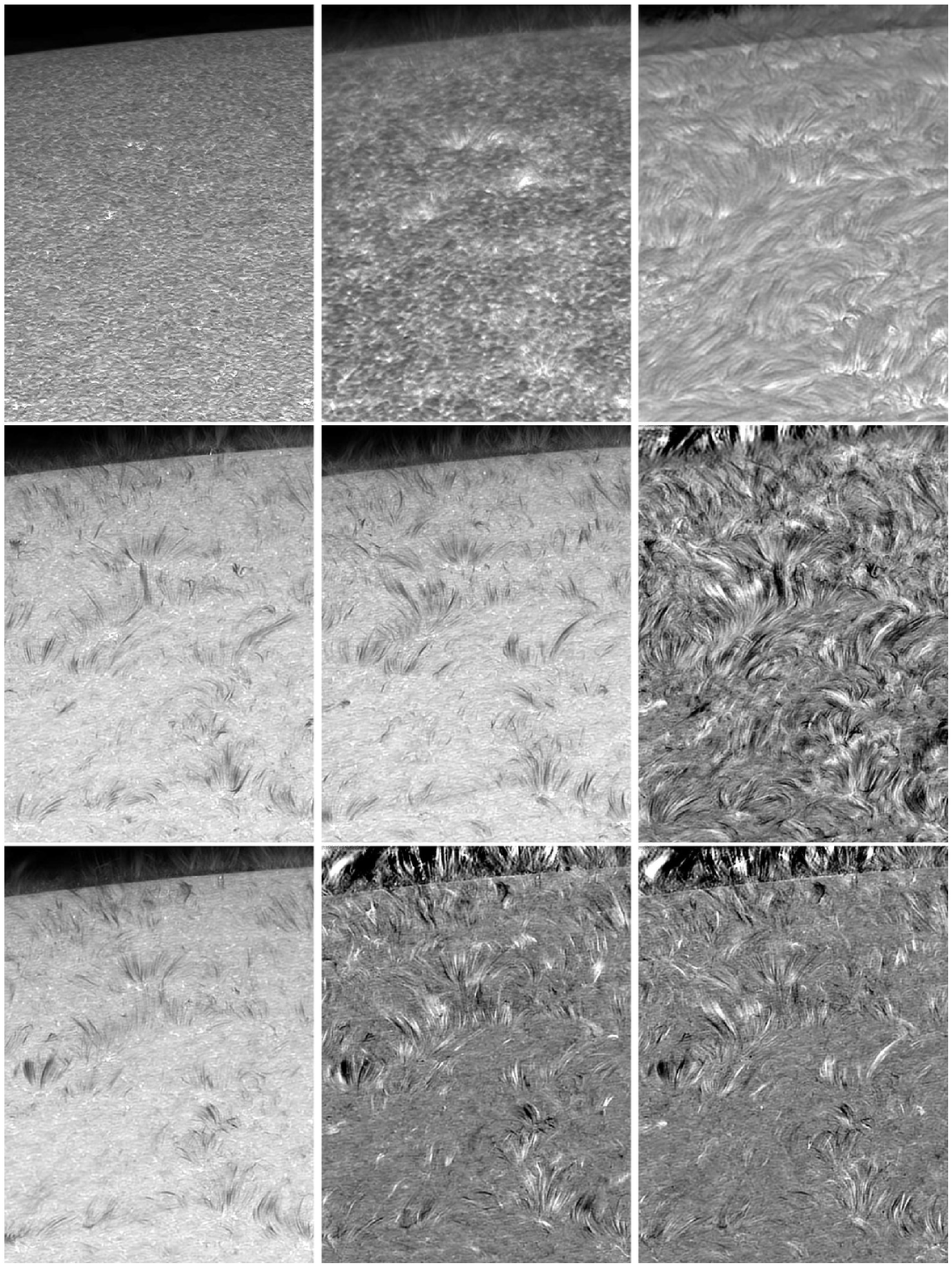}{fig:DOT-nine}{Near-limb
  DOT images taken on October 4, 2005 and available in the DOT
  database.  I suggest to zoom in with a PDF viewer to appreciate
  their detail. Field of view $60\arcsec \times 80\arcsec$.  Top row:
  G-band image showing photospheric granulation and faculae, \CaIIH\
  image showing bright network with thin straws and diffuse emission,
  elsewhere dark cell interiors with bright acoustic-shock grains,
  \Halpha\ line center image showing masses of fibrils, most of them
  covering cell interiors.  These three images were taken
  simultaneously.  Middle row: $\Delta \lambda \is -800$\,m\AA\ and
  $+800$\,m\AA\ images taken slightly earlier.  These show only
  upright fibrils jutting out from network. They are similar but not
  identical. Third panel: $\pm 600$\,m\AA\ time-delay Dopplergram at
  $\Delta t \is 1$~minute.  Bottom row: $-800$\,m\AA\ image taken four
  minutes earlier and $\pm 800$\,m\AA\ time-delay Dopplergrams at
  $\Delta t \is 2$ and 4 minutes.  Observation and alignment:
  P. S\"utterlin.  }

\articlefigure[width=10cm]{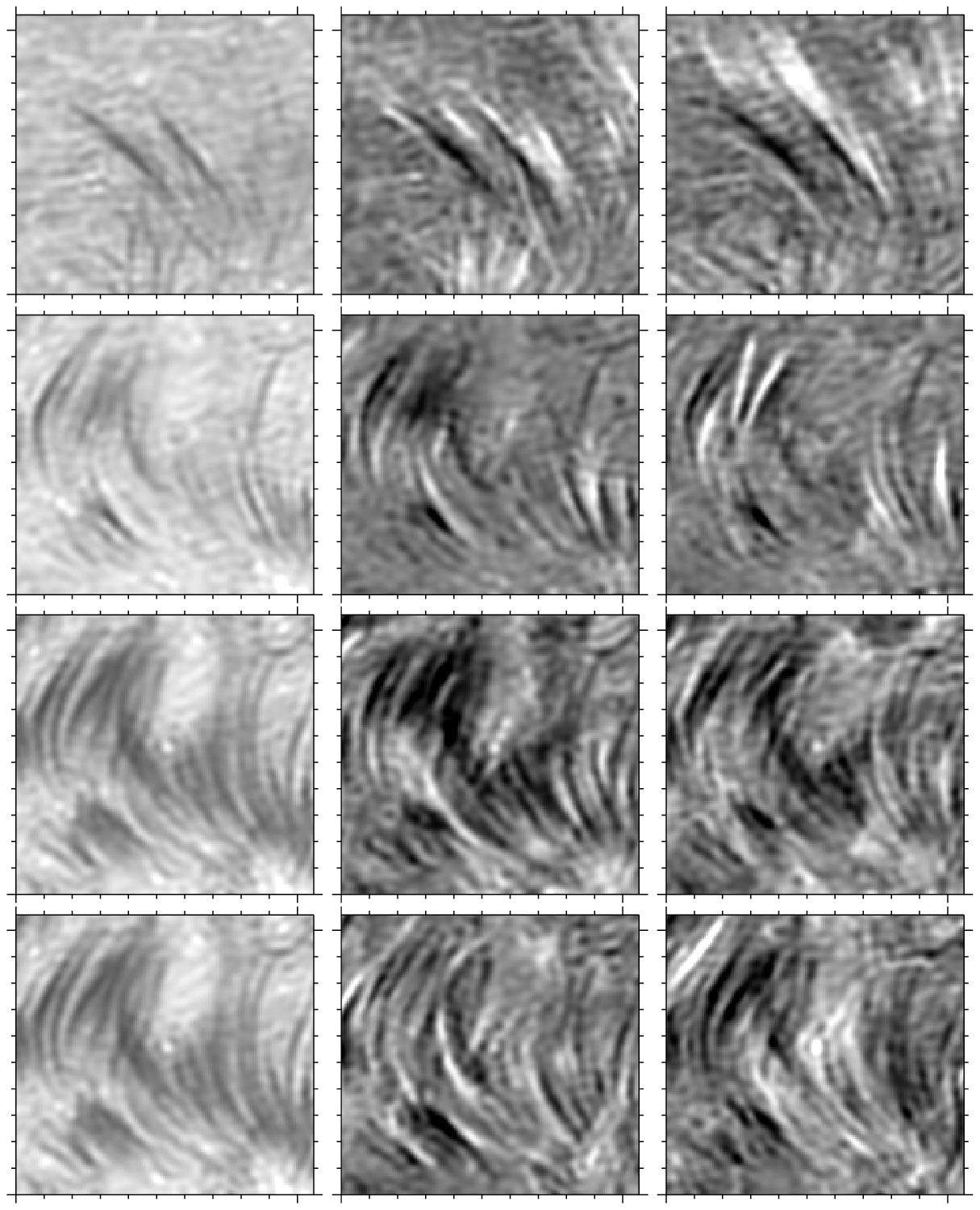}{fig:DOT-cuts}{First
  two rows: enlarged cutouts of the bottom panels in
  Fig.~\ref{fig:DOT-nine}, for subfields at the upper-right from the
  center and in the lower-right corner.  Third row: same subfield as
  in the second row, but using $\Delta \lambda \is \pm 600$\,m\AA\
  images (taken at slightly different times from those in the second
  row).  Bottom row: same subfield as in the second row, but the
  second and third panels are now same-wavelength time-delay
  difference images subtracting successive $-600$\,m\AA\ images at
  about two and four minutes delay.  The axis ticks are 1\arcsec\
  apart.}

\section{Old twist to spicules}

\altsubsubsection*{DOT near-limb \Halpha}
While preparing this presentation I realized, belatedly, that the old
DOT observations might still serve to illustrate these various type-II
spicule motions by constructing time-delay Dopplergrams
$[I_b(0)-I_r(\Delta t)]\,/\,[I_b(0)+I_r(\Delta t)]$ from a blue-wing
image $I_b$ and a red-wing image $I_r$ taken $\Delta t$ later
(Figs.~\ref{fig:DOT-nine} and \ref{fig:DOT-cuts}).

The first row of Fig.~\ref{fig:DOT-nine} displays the photospheric
scene and the overlying \Halpha\ line-center fibril carpet\footnote{
  The sharp limb is probably parasitic continuum light (\cf\
  \citeads{1974soch.book.....B}).}. 
The second row illustrates the similarity but non-identity of blue and
red outer-wing images at $\Delta \lambda \is \pm 800$\,m\AA.  One sees
the same bushes but made up of different spicular features. The third
panel is a time-delay Dopplergram at $\Delta \lambda \is \pm
600$\,m\AA\ and $\Delta t \is 1$~minute.  Bright features represent
spicules in the red-wing image.  The smaller wavelength separation is
chosen because for extreme Dopplershift the line-of-sight components
of outflow, to-from swaying, and torsional motion must add up
constructively and this is unlikely to happen in both wings at nearly
the same time.  At his smaller Dopplershift there are many instances
of aligned look-alike black and white features.  Thus, parts of a
given spicule bush may appear black or white, \ie\ Dopplershifted into
the blue or the red wing, without slant discrimination.  The third row
shows an earlier $-800$\,m\AA\ image and time-delay Dopplergrams with
it at two and four minutes delay.  Juxtaposition of dark and
subsequent bright features indicates opposite blueshift and redshift
at these time intervals.  At these longer delays these Dopplergrams
also show similar morphology of network spicule bushes in black and in
white in many locations.  Significant change occurs between the two
delay samplings.

Figure~\ref{fig:DOT-cuts} magnifies two subfields from the bottom row
of Fig.~\ref{fig:DOT-nine} and adds 600\,m\AA\ time-delay Dopplergrams
and same-wavelength subtractions for the second subfield.  Again the
overall black-and-white Dopplergram similarity is striking.  There are
good examples of aligned black-and-white feature pairs in the
Dopplergrams that indicate twist motion.  In the top row the bright
splash in the third panel extends out from the shorter feature seen
earlier, indicating rising outflow away from the viewer.  The pair of
brightest streaks in the third panel of the second row are adjacent to
similar but dark spicules in the blue-wing image, suggesting torsion
plus sway Doppler reversal.  The corresponding 600\,m\AA\ Dopplergram
underneath shows them only weakly, confirming that their redshifts are
extreme.

Transverse spicule swaying observed as proper motion (movement in the
plane of the sky) is better seen in higher-cadence movies as those of
\citetads{2007Sci...318.1574D}, 
but it is also clear in the difference images in the bottom row of
Fig.~\ref{fig:DOT-cuts}.  Adjacent look-alike black and white features
in these imply spicule swaying transverse to the line of sight.  It is
clearly evident in the first difference image, less in the second due
to spicule evolution.

Overall, the morphology of the bright features in these various
time-delay maps indeed suggests that many, if not most, spicules
combine motions in these various modes.  Clearly, high spatial
resolution, fast cadence, and good spectral resolution as well as
large spatial extent, long temporal duration, and multi-line spectral
sampling are the way to nail their role.  Also in simulations.

\altsubsubsection*{Last Utrecht eclipse expedition}
I then realized, yet more belatedly, that the tilt that the torsional
wave mode gives to spectral lines at the limb explains the old riddle
with which Houtgast and Namba concluded the Utrecht eclipse efforts
long ago.  At the 1973 eclipse
they used a large grazing-incidence slitless \'echelle spectrometer with an
image intensifier to take high-dispersion spectrograms in rapid
cadence at second and third contact.  The instrument and preparations
are described in detail in
\citetads{1979RNAAS..82..209H}, 
the observations in \citetads{1979RNAAS..82..223H}
-- 
for each author the last research papers.  The main result was that
the absorption lines showed varying tilts with respect to the emission
lines just before second contact.  Figure~\ref{fig:eclips} shows a
sample.  The tilts differ between the two Bailey beads and between
different absorption lines.  The maximum displacement was 0.08~\AA,
corresponding to 5\,\kms\ if interpreted as Dopplershift which is
more likely than slitless spatial mapping (corresponding to 30\arcsec\
because the grazing incidence compressed the spatial dimension in the
dispersion direction by a factor 3.4).

Houtgast and Namba were not sure that the tilts were not instrumental
and indeed could produce line tilts in subsequent laboratory
experiments with the spectrograph that came from the Dove prism used
for image rotation, but this could not explain differential tilt
between absorption and emission lines nor tilt variations.  They
concluded that the tilts remained without obvious explanation.  I
remember that Houtgast made a careful inventory of tilt per line type
(element, atom/ion, strength, excitation energy) but I fear that his
work is lost.

\articlefigure{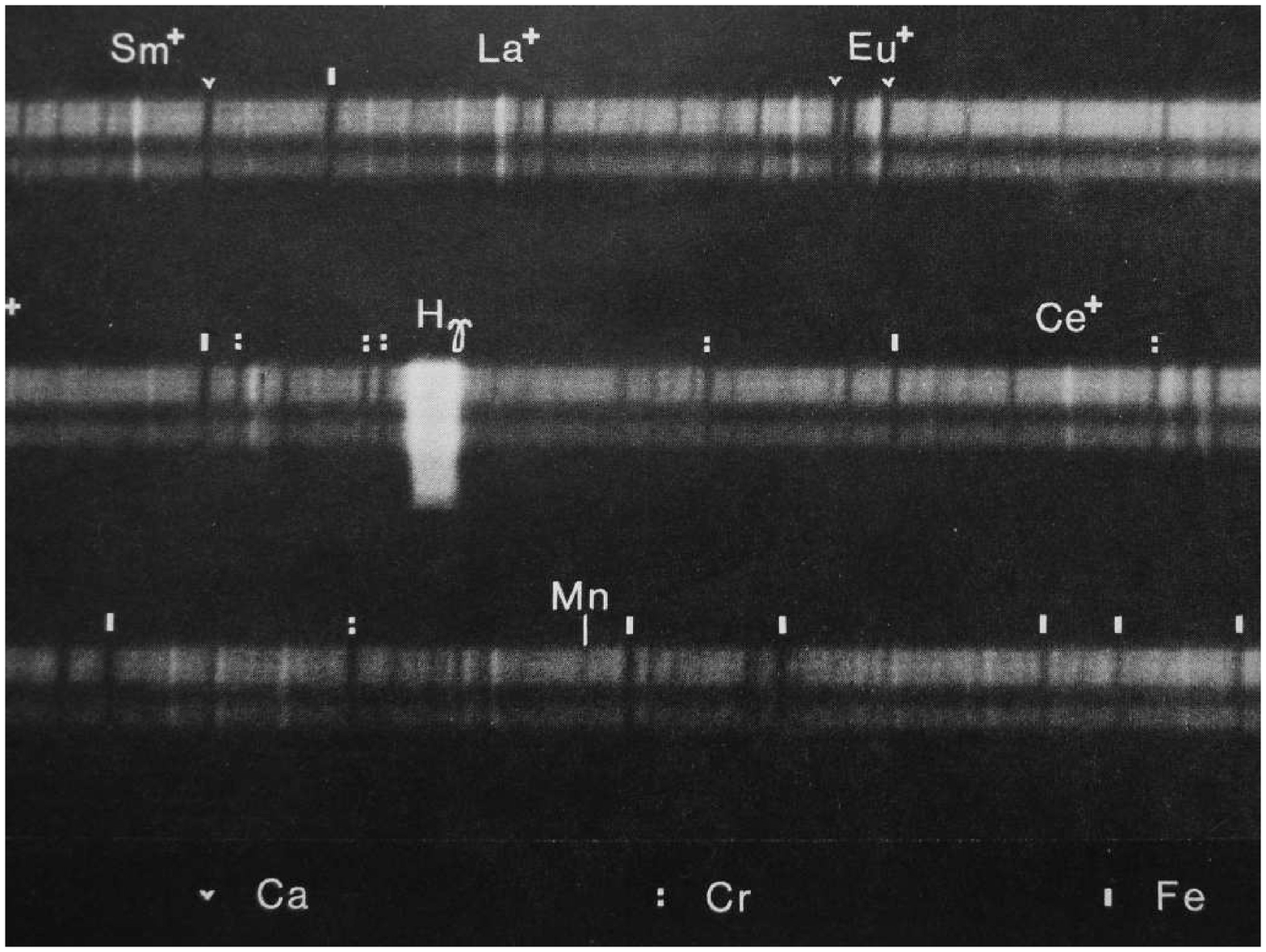}{fig:eclips}{Part of a
  spectrogram taken just before second contact of the June 30, 1973
  eclipse at Atar, Mauritania.  From
  \citetads{1979RNAAS..82..223H}. 
}

The answer appears simple now.  The emission lines are rare earth
lines which derive their emission from strong scattering with much
interlocking, as shown by
\citetads{1971A&A....10...64C} 
while he was a postdoc at Utrecht.  These lines do not show local
conditions but smoothed-out photospheric radiation from a wide area.
Instead, the absorption lines form close to LTE, respond to local
conditions, and likely sampled a single spicule or spicule bush as
selected by the Bailey bead (lunar limb valley), indeed with different
formation and Doppler response for different lines.  Their tilts then
formed comparably to the near-limb line tilts in Fig.~5 of
\citetads{2012arXiv1205.5006D}.  

\section{Discussion}
What are the next steps in spicule research?  Observationally NASA's
upcoming IRIS solar spectrometry mission (\url{http://iris.lmsal.com})
is slated to diagnose type-II spicules especially in \MgII\ \hk\ on
the disk.  In fast-scan imaging spectroscopy in these lines they are
likely to appear as thin strands that are Dopplershifted well out of
the central emission peaks to appear bright against the dark
internetwork background, made extra dark through coherent inner-wing
scattering.  Their transverse torsional and swaying motions will
contribute to this Doppler isolation.

In simulations the quest is to identify their acceleration and heating
mechanisms.  The fact that they occur in unipolar areas
(\citeads{2011ApJ...727....7M}) 
suggests that component reconnection rather than opposite-polarity
reconnection may be a major agent.  The simulation example of
\citetads{2011ApJ...736....9M} 
indeed points this way.

I look forward in particular to limb spectroscopy with IRIS.  More
than forty years ago Houtgast taught me to appreciate the extreme limb
in spectra.  It will be good to return to it exploiting the extended
seeing-free observing that space offers over eclipses.

\acknowledgements I thank many ex-SIU students and colleagues for
enriching my life with friendships and joy in joint research. It is a
large comfort that Utrecht-educated expats have strong roles in
ongoing chromosphere research including SST observing, the IRIS
mission, and numerical simulation.  It is not really a comfort that
the board of Utrecht University went this stupid only after my
mandatory retirement.


\begin{thebibliography}{}
\expandafter\ifx\csname natexlab\endcsname\relax\def\natexlab#1{#1}\fi
\expandafter\ifx\csname url\endcsname\relax
  \def\url#1{\texttt{#1}}\fi
\expandafter\ifx\csname urlprefix\endcsname\relax\def\urlprefix{URL }\fi
\providecommand{\eprint}[2][]{\url{#2}}

\bibitem[{{Beckers}(1964)}]{1964PhDT........83B}
{Beckers}, J.~M. 1964, Ph.D. thesis, Utrecht University, AFCRL Environmental
  Res.\ Paper 49

\bibitem[{{Beckers}(1968)}]{1968SoPh....3..367B}
--- 1968, \solphys, 3, 367

\bibitem[{{Beckers}(1972)}]{1972ARA&A..10...73B}
--- 1972, \araa, 10, 73

\bibitem[{{Bray} \& {Loughhead}(1974)}]{1974soch.book.....B}
{Bray}, R.~J., \& {Loughhead}, R.~E. 1974, {The solar chromosphere}, Chapman
  and Hall

\bibitem[{{Canfield}(1971)}]{1971A&A....10...64C}
{Canfield}, R.~C. 1971, \aap, 10, 64

\bibitem[{{De Pontieu} et~al.(2012){De Pontieu}, Carlsson, {Rouppe van
      der Voort}, Rutten, Hansteen, \&
    Watanabe}]{2012arXiv1205.5006D} {De Pontieu}, B., Carlsson, M.,
    {Rouppe van der Voort}, L. H.~M., Rutten, R.~J., Hansteen, V.~H.,
    \& Watanabe, H. 2012, ApJ, in press, preprint arXiv 1205.5006

\bibitem[{{De Pontieu} et~al.(2007{\natexlab{a}}){De Pontieu}, {Hansteen},
  {Rouppe van der Voort}, {van Noort}, \& {Carlsson}}]{2007ApJ...655..624D}
{De Pontieu}, B., {Hansteen}, V.~H., {Rouppe van der Voort}, L., {van Noort},
  M., \& {Carlsson}, M. 2007{\natexlab{a}}, \apj, 655, 624

\bibitem[{{De Pontieu} et~al.(2007{\natexlab{b}}){De Pontieu},
    {McIntosh}, {Hansteen}, {Carlsson}, {Schrijver}, {Tarbell},
    {Title}, {Shine}, {Suematsu}, {Tsuneta}, {Katsukawa}, {Ichimoto},
    {Shimizu}, \& {Nagata}}]{2007PASJ...59S.655D} {De Pontieu}, B.,
  {McIntosh}, S., {Hansteen}, V.~H., {Carlsson}, M., {Schrijver},
  C.~J., {Tarbell}, T.~D., {Title}, A.~M., {Shine}, R.~A., {Suematsu},
  Y., {Tsuneta}, S., {Katsukawa}, Y., {Ichimoto}, K., {Shimizu}, T.,
  \& {Nagata}, S. 2007{\natexlab{b}}, \pasj, 59, 655

\bibitem[{{De Pontieu} et~al.(2011){De Pontieu}, {McIntosh}, {Carlsson},
  {Hansteen}, {Tarbell}, {Boerner}, {Mart{\'{\i}}nez-Sykora}, {Schrijver}, \&
  {Title}}]{2011Sci...331...55D}
{De Pontieu}, B., {McIntosh}, S.~W., {Carlsson}, M., {Hansteen}, V.~H.,
  {Tarbell}, T.~D., {Boerner}, P., {Mart{\'{\i}}nez-Sykora}, J., {Schrijver},
  C.~J., \& {Title}, A.~M. 2011, Science, 331, 55

\bibitem[{{De Pontieu} et~al.(2007{\natexlab{c}}){De Pontieu}, {McIntosh},
  {Carlsson}, {Hansteen}, {Tarbell}, {Schrijver}, {Title}, {Shine}, {Tsuneta},
  {Katsukawa}, {Ichimoto}, {Suematsu}, {Shimizu}, \&
  {Nagata}}]{2007Sci...318.1574D}
{De Pontieu}, B., {McIntosh}, S.~W., {Carlsson}, M., {Hansteen}, V.~H.,
  {Tarbell}, T.~D., {Schrijver}, C.~J., {Title}, A.~M., {Shine}, R.~A.,
  {Tsuneta}, S., {Katsukawa}, Y., {Ichimoto}, K., {Suematsu}, Y., {Shimizu},
  T., \& {Nagata}, S. 2007{\natexlab{c}}, Science, 318, 1574

\bibitem[{{De Pontieu} et~al.(2009){De Pontieu}, {McIntosh}, {Hansteen}, \&
  {Schrijver}}]{2009ApJ...701L...1D}
{De Pontieu}, B., {McIntosh}, S.~W., {Hansteen}, V.~H., \& {Schrijver}, C.~J.
  2009, \apjl, 701, L1

\bibitem[{{Hansteen} et~al.(2006){Hansteen}, {De Pontieu}, {Rouppe van der
  Voort}, {van Noort}, \& {Carlsson}}]{2006ApJ...647L..73H}
{Hansteen}, V.~H., {De Pontieu}, B., {Rouppe van der Voort}, L., {van Noort},
  M., \& {Carlsson}, M. 2006, \apjl, 647, L73

\bibitem[{{Houtgast}(1942)}]{1942QB551.H68......}
{Houtgast}, J. 1942, Ph.D. thesis, Utrecht University

\bibitem[{{Houtgast} \& {Namba}(1979{\natexlab{a}})}]{1979RNAAS..82..209H}
{Houtgast}, J., \& {Namba}, O. 1979{\natexlab{a}}, Procs.\ Royal Netherlands
  Acad.\ Arts and Sciences, 82, 209

\bibitem[{{Houtgast} \& {Namba}(1979{\natexlab{b}})}]{1979RNAAS..82..223H}
--- 1979{\natexlab{b}}, Procs.\ Royal Netherlands Acad.\ Arts and Sciences, 82,
  223

\bibitem[{{Leenaarts} et~al.(2012){Leenaarts}, {Carlsson}, \& {Rouppe van der
  Voort}}]{2012ApJ...749..136L}
{Leenaarts}, J., {Carlsson}, M., \& {Rouppe van der Voort}, L. 2012, \apj, 749,
  136

\bibitem[{{Mart{\'{\i}}nez-Sykora} et~al.(2011){Mart{\'{\i}}nez-Sykora},
  {Hansteen}, \& {Moreno-Insertis}}]{2011ApJ...736....9M}
{Mart{\'{\i}}nez-Sykora}, J., {Hansteen}, V., \& {Moreno-Insertis}, F. 2011,
  \apj, 736, 9

\bibitem[{{McIntosh} et~al.(2011){McIntosh}, {Leamon}, \& {De
  Pontieu}}]{2011ApJ...727....7M}
{McIntosh}, S.~W., {Leamon}, R.~J., \& {De Pontieu}, B. 2011, \apj, 727, 7

\bibitem[{{Mihalas}(1970)}]{1970stat.book.....M}
{Mihalas}, D. 1970, {Stellar atmospheres}, 1st edition, Freeman, San
  Francsico

\bibitem[{{Mihalas}(1978)}]{1978stat.book.....M}
--- 1978, {Stellar atmospheres}, 2nd edition, Freeman, San Francsico

\bibitem[{{Rouppe van der Voort} et~al.(2009){Rouppe van der Voort},
  {Leenaarts}, {De Pontieu}, {Carlsson}, \& {Vissers}}]{2009ApJ...705..272R}
{Rouppe van der Voort}, L., {Leenaarts}, J., {De Pontieu}, B., {Carlsson}, M.,
  \& {Vissers}, G. 2009, \apj, 705, 272

\bibitem[{{Rutten}(2006)}]{2006ASPC..354..276R}
{Rutten}, R.~J. 2006, in Solar MHD Theory and Observations: A High Spatial
  Resolution Perspective, edited by {J.~Leibacher, R.~F.~Stein, \&
  H.~Uitenbroek}, \aspcs, 354, 276

\bibitem[{{Rutten}(2007)}]{2007ASPC..368...27R}
--- 2007, in The Physics of Chromospheric Plasmas, edited by {P.~Heinzel,
  I.~Dorotovi{\v c}, \& R.~J.~Rutten}, \aspcs, 368, 27

\bibitem[{Rutten(2012)}]{2011arXiv1110.6606R}
Rutten, R.~J. 2012, Phil.\ Trans.\ Royal Soc., in press, preprint ArXiv
  1110.6606

\bibitem[{{Rutten} \& {Uitenbroek}(2012)}]{2012A&A...540A..86R}
{Rutten}, R.~J., \& {Uitenbroek}, H. 2012, \aap, 540, A86

\bibitem[{{Suematsu} et~al.(2008){Suematsu}, {Ichimoto}, {Katsukawa},
  {Shimizu}, {Okamoto}, {Tsuneta}, {Tarbell}, \& {Shine}}]{2008ASPC..397...27S}
{Suematsu}, Y., {Ichimoto}, K., {Katsukawa}, Y., {Shimizu}, T., {Okamoto}, T.,
  {Tsuneta}, S., {Tarbell}, T., \& {Shine}, R.~A. 2008, in First Results From
  Hinode, edited by {S.~A.~Matthews, J.~M.~Davis, \& L.~K.~Harra}, \aspcs, 
  397, 27

\bibitem[{{Tian} et~al.(2011){Tian}, {McIntosh}, {De Pontieu},
  {Mart{\'{\i}}nez-Sykora}, {Sechler}, \& {Wang}}]{2011ApJ...738...18T}
{Tian}, H., {McIntosh}, S.~W., {De Pontieu}, B., {Mart{\'{\i}}nez-Sykora}, J.,
  {Sechler}, M., \& {Wang}, X. 2011, \apj, 738, 18

\end{thebibliography}

\end{document}